This paper has been accepted for publication and is forthcoming in *The Global and Digital Governance Handbook.*

## A Multilevel Framework for AI Governance

Hyesun Choung, Ph.D.
Michigan State University

Prabu David, Ph.D.
Michigan State University

John S. Seberger, Ph.D.
Drexel University

**Abstract**

To realize the potential benefits and mitigate potential risks of AI, it is necessary to develop a framework of governance that conforms to ethics and fundamental human values. Although several organizations have issued guidelines and ethical frameworks for trustworthy AI, without a mediating governance structure, these ethical principles will not translate into practice. In this paper, we propose a multilevel governance approach that involves three groups of interdependent stakeholders: governments, corporations, and citizens. We examine their interrelationships through dimensions of trust, such as competence, integrity, and benevolence. The levels of governance combined with the dimensions of trust in AI provide practical insights that can be used to further enhance user experiences and inform public policy related to AI.

**Chapter Keywords**

Artificial Intelligence, Governance, Trust, Ethics, Multilevel Framework

**Key Discussion Areas**

- Relevance and acceptance of AI by its key stakeholders
- Governance models for ensuring trustworthy AI
- Offering a multidimensional, multilevel model of trust in AI
- Corporate self-governance and government regulation of AI

## 1.      Introduction

Artificial intelligence (AI) systems are becoming increasingly adept in their ability to learn, reason, self-correct, and emulate human decisions in various domains (Russell et al. 2016; Turing 1950; Watson 2019).  Modern AI systems are driven by machine learning (ML) techniques to accomplish tasks such as recognizing faces and voices, reading X-rays, screening job applicants, identifying credit card fraud, and enabling law enforcement. Alongside professional and institutional use of ML-driven AI, consumer applications of AI are increasingly ubiquitous in daily personal life—embedded in home devices, smart city management, and autonomous driving systems (Gorwa, Binns, and Katzenbach 2020; Lockey et al. 2021). AI is also used to tackle challenging social problems, including the maintenance of public health (Wahl et al. 2018), social justice (Graham and Hopkins 2021), and public safety (Kankanhalli et al. 2019).

Even though governments, corporations, and individuals benefit from the successes of AI technologies, the detrimental effects of AI are increasingly apparent, including algorithmic biases and challenges to civil liberties, increased surveillance (Whittaker et al. 2018), and the diminution of human agency (Seberger 2021). To realize the potential benefits of AI, it is necessary to address the detrimental effects and risks of AI. Such a process is further complicated by AI's reliance on ML algorithms that have been characterized as a black box, because the inner workings of these algorithms are opaque and not easily explainable (Barredo Arrieta et al. 2020). Though some researchers have developed methods to translate the deep learning inside the black box into formal rules, this area is still nascent (Samek et al. 2019)—hence the legitimately pressing concerns about the lack of transparency and understanding in current AI technologies (Shin and Park 2019). The pervasiveness of AI and the autonomous decisions it renders in many areas of life, compounded by its black-box approach to sensitive data, have created an urgent need for well-defined and human-centered principles of governance that conform to ethics and fundamental human values (Jobin, Ienca, and Vayena 2019). Such principles of governance may mitigate the risks of AI and make it possible to realize AI's full potential.

In the past few years, influential groups and organizations have addressed the need for governance by issuing guidelines and ethical frameworks for the development and deployment of trustworthy AI. These include the Organization for Economic Co-operation and Development (OECD 2019), the European Commission's High Level Expert Group on Artificial Intelligence (AI HLEG 2019), the Institute of Electrical and Electronics Engineers (IEEE), Microsoft, DeepMind, and Google (Hagendorff 2020; Jobin, Ienca, and Vayena 2019), organizations driven by different motives, ethical imperatives, and missions.

However, in the absence of a mediating governance structure, ethical principles do not automatically translate into practice (Mittelstadt 2019). Given the wide reach of AI across social, cultural, and national borders, governance structures for AI require guiding principles sensitive to the contexts in which they will be deployed. For example, the concept of privacy has widely variable definitions across cultures (Farrall 2008). A codified principle for ethical AI that involves privacy would, therefore, need to be sensitive to such cultural variance. While individual rights and autonomy are foundational in the AI HLEG (2019) ethical guidelines, such Western constructs may not map cleanly or ethically to social contexts where individual rights are subsidiary to the rights of the political party and the state. Despite the complexity of intercultural differences, three groups of dynamically interdependent stakeholders are universally implicated in the concern for ethical AI governance: governments, corporations, and citizens. Any baseline for AI governance, then, necessarily involves these three groups and their interrelationships. We approach such interrelationships through the lens of trust.

## 2.      A Multilevel Approach to AI Governance for Trust-Enhancing Practices

In the broadest sense, trust is defined as a confident relationship with the unknown (Botsman 2017). As we—users, scholars, designers, and policy-makers—prepare for a future with ubiquitous AI and its unknown dimensions, trust is essential to realizing a future in which the risks associated with machine intelligence are mitigated and its benefits are nurtured. Effective governance is a cornerstone of trust and is necessary to mediate the relationship between humans and AI technologies. A framework for governance of AI can be achieved only by acknowledging the agency of governments, organizations, and citizens, and their interdependencies.

In the following sections, we review governance principles from three perspectives: the transnational/national (i.e., governmental), the organizational (i.e., corporate), and the individual. We provide a detailed review of trust at the individual level from a psychological perspective. This review is followed by an examination of ethical frameworks for governance that have been adopted by the European Union (EU), IEEE, and Big Tech. We conclude by offering a multilevel framework of AI governance that is built on trust and guided by ethics.

3.      **International and National Governance**

Amid the fierce competition among countries for global leadership in AI (Castro 2019), some efforts toward international consensus building have been undertaken. The International Panel on Artificial Intelligence (IPAI) at the Group of Seven (G7) summit in 2019 aimed to follow the example of the International Panel on Climate Change (IPCC) to support the responsible development of AI "grounded in human rights, inclusion, diversity, and innovation" (Government of Canada 2019). In 2020, the Group of Twenty (G20) countries committed to advance the G20 AI Principles drawn from the OECD's recommendations about AI. These principles seek to foster public trust and confidence in AI by promoting values such as inclusiveness, human-centricity, transparency, robustness, and accountability (Box 2020).

Given AI's potential to improve health, education, agriculture, manufacturing, and energy, countries and international bodies are unsurprisingly eager to develop and adopt AI. At the same time, such eagerness is countered by the perceived threat of machine agency and automated decision-making, as well as the displacement of humans by automation and robots. Such risks associated with AI are perceived differently in different societies relative to dominant political ideology, value systems, and culture norms. For example, citizens from traditional collectivist cultures, such as China, may be more inclined than citizens from individualist cultures to sacrifice personal data and choose security over privacy (Kostka, Steinacker, and Meckel 2021). Despite the unavoidable cultural and contextual differences, a mix of international principles and policies are urgently needed to build trustworthy and human-centered AI (OECD 2020).

Such efforts are underway. The Beijing Academy of Artificial Intelligence has released a set of principles endorsed by leading Chinese universities and organizations (Beijing AI Principles 2019), which includes basic human values such as the good of humanity, diversity and inclusion, and ethics as founding principles. Similar values were highlighted by the European Commission's AI HLEG, including human dignity and autonomy, prevention of harm, fairness, explicability, accountability, privacy, and social and environmental well-being (AI HLEG 2019). Despite the similarities, the differences in political systems in China and the EU will eventually shape the implementation of these values.

In the United States, despite efforts to pass AI governance policies in state legislatures (National Conference of State Legislatures 2021), there have been no major AI-specific legislative successes at the federal or state level. One related legislation, however, is the California Consumer Protection Act (CCPA), which focuses on data privacy and generally emulates the General Data Protection Regulation (GDPR) enacted by the EU (Somers and Boghaert 2018). Though not directly related to AI, the CCPA could have a significant effect on how the data practices of AI technologies become normalized.

In summary, international bodies and nations have proposed key principles of trustworthy AI. Many of these principles are founded on the values of human rights, such as dignity, freedom, and autonomy. Such principles optimistically envisage a future in which human intelligence coexists with artificial intelligence. However, enforcement or the real-world implications of these policies on technology firms, software developers, and data brokers is not clear. As described in the next section, technology firms and oversight bodies have taken it upon themselves to develop individual policies in the tradition of corporate self-governance.

### 4.      Corporate Self-Governance

AI is designed and deployed by corporations, who have their own role to play in governance. In addition to implementing national policy, industries and corporations can complement government regulations with self-governance. A notable advance is a call to action for ethically aligned design for business created by the IEEE Global Initiative on Ethics of Autonomous and Intelligent Systems (Chatila and Havens 2019). The IEEE's policy calls for participatory design and attention to ethics throughout the AI development cycle. Such guidance frames AI governance as a sociotechnical necessity that could positively impact humanity by underscoring the importance of trust, transparency, and accountability among corporate leaders.

Similar policies have been adopted by Big Tech companies. Corporations such as Google and Microsoft have developed policies, which are posted on their websites. Google's policy states that the company is optimistic about AI and is sensitive to its potential harm (Google, n.d.). Key values that inform Google's position include empowering humanity, uplifting society, and addressing fairness and bias. Concern for privacy and accountability are also mentioned: Google states that it will not use surveillance that violates internationally accepted norms, nor will it pursue weapons or systems intended principally to cause harm. Microsoft similarly emphasizes responsible and trustworthy AI that empowers others and encourages the use of AI for socially responsible outcomes (Microsoft, n.d.). The importance of trust is widely discussed in Microsoft's policy. Other companies like Apple, Amazon, and Facebook employ AI in their consumer-facing products and demonstrate a sensitivity to trust in written statements, which is integral to corporate reputation.  It appears that Big Tech companies are keenly aware of the importance of trust. Their approach to self-governance hinges on building trust through ethical practices promoted in their AI policies. It remains to be seen, however, how accurately such written statements reflect the real-world practices of such companies.

In the United States, tech corporations exercise significant influence on both technological development and regulatory methods (Cihon, Schuett, and Baum 2021). Although companies formulate their own ethical guidelines or adopt ethically motivated self-commitments, they currently do not have clear standards for practice or internal assessors and are hesitant to create a binding legal framework (Hagendorff 2020). One way to enhance corporate self-governance is to assess and mitigate risk from design to implementation and to institute a certification or accreditation program for AI developers who are trained to anticipate risks (Roski et al. 2021). Certification would also include training in ethics, and some have recommended virtue ethics as a promising approach (Hagendorff, 2020), which differs from traditional ethics approaches in that it locates ethics in specific contexts dependent upon traits of individuals rather than universal codes of conduct. Another suggestion is developing and initiating "ethics-based auditing" for organizations and academic researchers who develop or deploy AI (Mökander and Axente 2021).

Through self-governance and ethics in training and evaluation processes, corporations have the power to shape our shared computational futures. However, as in many areas, corporate self-governance can be held more accountable through higher-level policy from governments that emphasize common good over corporate gain. Much like the added accountability that accrues from top-down governmental policies, bottom-up demand from citizens can drive accountability in corporate governance.

### 5.      AI Literacy and Governance by Citizens

Citizens have a right to trustworthy AI. To empower individuals to claim agency, enhancing citizens' competencies and literacy in AI is essential. User empowerment requires interventions that offer an alternative to the digital resignation — a sociotechnical phenomenon resembling learned helplessness in which users resign themselves to having little agency over how their data is used — that some individuals have grown to accept (Seberger et al. 2021). It begins with providing users a meaningful understanding of a broad set of sociotechnical factors that contribute to and characterize their use of an AI technology. AI literacy has been defined as "a set of competencies that enables individuals to critically evaluate AI technologies; communicate and collaborate effectively with AI; and use AI as a tool online, at home, and in the workplace" (Long and Magerko 2020, 598). Recent efforts to foster AI literacy in the United States include an open learning opportunity for underrepresented students (AI4ALL 2019) and integration of AI programs into K-12 curricula (Zimmerman 2018).

Given the rapid pace at which AI is evolving, however, it is unclear how prior AI literacy campaigns remain relevant to the changing landscape of AI. For example, recently, ChatGPT developed by OpenAI has taken the world by storm because of its natural language processing ability and fluency in writing. While the responses provided by the ChatGPT may sound convincing, until the technology matures, literacy interventions may be required to train citizens to critically evaluate the quality and veracity of responses.

However, AI is a relatively new technology, and the majority of citizens hold optimistic perspectives about AI without a sophisticated knowledge of potential risks (Carrasco et al. 2019). Relying purely on citizen action for an emerging technology whose future is unknown is unrealistic and even unfair. In the long term, such an approach may amount to negligence. While government- and corporate-led efforts to inform and educate the public are essential, an integrated approach that includes citizens, corporations, and government is likely the most responsible and robust approach.

## 6.    Psychology of Trust

To develop a model of governance for a new and potentially disruptive technology that touches the lives of citizens, trust is essential (Shin 2021; Wu et al. 2011). But the beginning of the 21st century has been marked by a significant erosion of trust in institutions and governments (Edelman 2021). The logical response to this crisis of trust is to build systems that instill confidence among citizens from different cultural backgrounds and values systems (Gillath et al. 2021; Thiebes, Lins, and Sunyaev 2020).

Although earlier we defined trust as "a confident relationship with the unknown" (Botsman 2017), here we examine trust in depth as a cornerstone of all relationships. Among individuals, trust is defined as "a psychological state comprising the intention to accept vulnerability based upon positive expectations of the intentions or behavior of another" (Rousseau et al. 1998, 395). Besides vulnerability and positive expectations of one another, trust combines characteristics, intentions, and behaviors (J. D. Lee and See 2004; Mayer, Davis, and Schoorman 1995) and is required to build mutuality and interdependence between parties.

Trust in humans can be defined as an amalgam of one's belief in another's ability, benevolence, and integrity (Mayer, Davis, and Schoorman, 1995). Ability refers to skills and competencies to successfully complete a given task. Benevolence pertains to whether the trusted individual or party has positive intentions that are not based purely on self-interest. And integrity points to the sense of morality and justice of the trusted party and is related to attributes such as consistency, predictability, and honesty. Absent such characteristics, meaningful interpersonal interactions cannot be forged.

Though interpersonal trust has been applied to technology with human-like characteristics (Calhoun et al., 2019; Gillath et al. 2021), some researchers caution that the principles of trust among humans cannot be applied directly to human-to-machine trust (Madhavan and Wiegmann 2007). Researchers have noted that trust in technology is qualitatively different from trust in people (McKnight et al. 2011). The key distinction is that a human is a moral agent, whereas technology lacks volition and moral agency. However, given the extent to which AI technologies manifest the value system and agency of the designer, this may no longer be a tenable distinction. Beginning with the dimensions of trust in human interaction, McKnight and colleagues (2011) revised the three dimensions of trust in technology as functionality, reliability, and helpfulness. Functionality refers to the capability of the technology, which the authors likened to human ability. Reliability is the consistency of operation, which is analogous to integrity. Helpfulness indicates whether a specific technology is useful to users and maps onto the benevolence dimension of human trust.

But AI is more than just a technology—it is an emergent sociotechnical system that strives for autonomy by replacing tasks and decisions made by humans. In that sense, McKnight's definition and the dimensions of trust in technology need to be adapted to the AI context. Unlike prior information technologies that relied on user input and the execution of rules programmed by humans, AI-driven technologies are capable of learning on their own and exercising functional autonomy, such as making decisions. AI's ostensible "intelligence" can be traced to its contextual awareness, data collection, processing, and decision-making abilities.

The autonomy and decision-making of AI in social domains, such as interpreting language, recognizing faces, or screening job applications, have led experts to sound the alarm on potential risks. Lankton, McKnight, and Tripp (2015) found that, when humanness is embedded or perceived in the technology, a trust-in-humans scale works better than a trust-in-technology scale. And when these technologies are depicted as capable of human qualities, including reasoning and motivations, they can induce high expectations and initial trust (Glikson and Woolley 2020). These findings suggest that trust in the human dimension of technology is a dynamic concept that varies considerably based on the context and the characteristics of the trusted agent. In short, trust in AI includes components of trust in people and components of trust in technology (Choung, David, and Ross 2022b).

Trust in automation (J. D. Lee and See 2004) is another framework relevant to the conceptualization of trust in AI and is anchored on performance, process, and purpose. Performance in automation refers to operational characteristics, including its reliability and ability. Process corresponds to the consistency of behaviors. Purpose describes why the automation was developed and the designers' intent. Table 6.1 summarizes the dimensions of trust in people, technology, and automation.

*Table 6.1 Dimensions of Trust*

| **Basis of Trust in People**<br>Mayer, Davis, and Schoorman (1995) | **Basis of Trust in Technology**<br>Lankton, McKnight, and Tripp (2015) | **Basis of Trust in Automation**<br>Lee and Moray (1992) |
|---|---|---|
| Competence/ability | Functionality | Performance |
| Integrity | Reliability | Process |
| Benevolence | Helpfulness | Purpose |

Multiple approaches to framing trust in technologies have been developed because our propensity to trust machines is interesting in and of itself. Researchers have identified contexts in which humans are more willing to entrust personal information to machines than to other humans, which has been explained as the machine heuristic (Sundar and Kim 2019). This is a robust effect that can be elicited by adding simple interface cues that suggest that the user is interacting with a machine. In turn, these cues are enough to prime machine characteristics such as accuracy, objectivity, neutrality, and unbiasedness. While some of these characteristics may be well deserved, some of this trust may be misplaced because machine learning algorithms that rely on data that represent human behaviors and practices may be inadvertently perpetuating human biases in AI.

Positive appraisals of machines also extend to algorithmic appreciation (Logg, Minson, and Moore 2019), the belief that algorithms are smarter, more objective, better decision-makers than humans. The countervailing perception to algorithmic appreciation is algorithmic aversion, the tendency to be critical of algorithms for their reductionism, lack of humanity, and subjective thinking (Dietvorst, Simmons, and Massey 2015, 2018). The psychology of machine heuristics, algorithmic appreciation, and algorithmic aversion justify the need for a better understanding of the relational attributes of trust between humans and machines. As governance policies for AI are designed, one must pay heed to warnings that the human tendency to treat computers as social actors is changing rapidly given our growing experience with computers and AI (Gambino, Fox, and Ratan 2020). We must develop a suitable understanding of trust that is sensitive to the emerging relationships between humans and AI.

## 7.    Propensity to Trust

In the previous section, we focused on the dimensions of trust. Now we return to the multilevel conceptualization of governance with individuals, corporations, and governments as stakeholders and the trust propensity at each level. The multilevel conceptualization combined with trust propensities accommodates interdependencies and power differences among the three stakeholder groups, resulting in a governance model that spans from the individual (i.e., intrapersonal and interpersonal) to the collective (i.e., institutions and society) (Fulmer and Dirks 2018). Such a multilevel framework of trust furthers our understanding of the process through which trust evolves and is crystallized over time (Hoff and Bashir 2015).

Trust at the individual level begins as a disposition or a general tendency to trust another person (Mayer, Davis, and Schoorman1995). This is known as propensity of trust and is predictive of initial trustworthiness (Alarcon et al. 2018; Colquitt, Scott, and LePine 2007). Much like the propensity to trust people, propensities to trust corporations, institutions, industries, and governments must be considered when building AI systems that aspire to trustworthiness. For example, an individual with a high trust propensity for other individuals may have a low trust in Facebook and support social media regulation but support development of AI for social good. A multilevel, multidimensional understanding of trust would allow for a governance design that can be principled and yet flexible to accommodate such nuances and complexity.

## 8.     Ethics and Trust Lenses in the Multilevel Framework

Corporations (e.g., Google, Microsoft) and national and international bodies (e.g., IEEE, European Commission) have identified ethical values and requirements that include the following themes: privacy protection, fairness, diversity, nondiscrimination and social justice, accountability, robustness, safety, resilience, transparency, explainability, human autonomy, loss of human jobs, need for human oversight, and limiting the use of AI as a weapon in wars (Hagendorff 2020). While there is significant overlap in the values and themes emphasized by different groups, the policy from the European Commission's AI HLEG (2019) offers a framework that is sufficiently nuanced and is an inspiring call to action. The AI HLEG framework rests on four ethical values—human autonomy, prevention of harm, fairness, and explicability. In addition, it underscores the rights of vulnerable members of society and the historically marginalized. The authors recognize the potential benefits and risks of AI and the need for risk mitigation. Drawing from these values and ethical principles, seven requirements are offered (see table 8.1).

*Table 8.1 Seven Ethics Requirements for Trustworthy AI Proposed by the European Commission's High Level Expert Group on Artificial Intelligence*

| Ethics Requirements for Trustworthy AI | Description |
| --- | --- |
| Human agency and oversight | AI systems should allow people to make informed decisions. There should be a human oversight mechanism through a "human-in-the-loop" approach. |
| Technical robustness and safety | AI systems should be safe, reliable, and reproducible to minimize unintended harm. |
| Privacy and data governance | Ensure privacy and data protection, which requires adequate data governance framework. |
| Transparency | AI systems and business models should be transparent, and the AI systems' decisions should be explainable to the stakeholders. People need to be informed about the systems' capabilities and limitations. |
| Diversity, nondiscrimination, and fairness | AI systems should be accessible to all, and unfair biases should be avoided. Minimizing algorithmic bias is also important. |
| Societal and environmental well-being | AI systems should benefit human beings and they should take into account the social impact and environmental consequences. |
| Accountability | AI mechanisms should be put in place to ensure responsibility and accountability for AI systems and their outcomes. |

These requirements contribute to the three trust dimensions of people: competence, integrity, and benevolence. These requirements of trust can help to identify possible foundations of trustworthy AI (Choung, David, and Ross, 2022a). For example, the competence dimension encompasses aspects of how an AI system functions, taking into consideration such aspects as safety, robustness, accountability, and explainability. The integrity dimension focuses on such human-centered characteristics of trust as fairness, nondiscrimination, privacy, and transparency. In concert, integrity and competence serve as guardrails for the sociotechnical configuration of AI. While these two dimensions may be sufficient to build trustworthy AI in the short term, the future of AI as a cohabitant of the human ecosystem requires enlightened approaches to computing in which machines must be trained to be benevolent. The best part of our humanity is in our willingness to look past individual self-interest and to make commitments and sacrifices for the common good. Creative approaches to nurture such benevolent values like respect for human autonomy, social justice, social and environmental well-being, and compassionate computing are emerging areas that should be considered as an ethical lens for trustworthy AI of the future.

*Figure 8.1 The Framework of Multilevel Governance and Trust in AI*

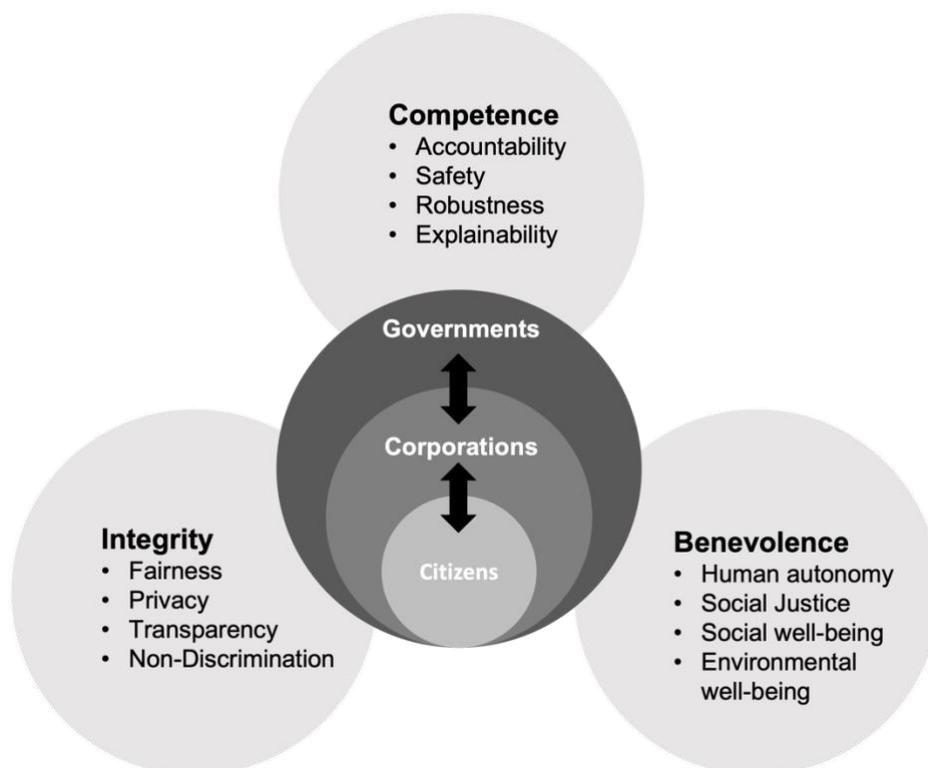

In the multilevel framework we outline in figure 8.1, each level—governmental, corporate, and individual—serves as a lens through which to interrogate the impacts of AI on the complex assemblage of people, corporations, and governments. More specifically, through such lenses we might systematically observe and test the effects of different attempts at trustworthiness in AI, ultimately arriving at an understanding of an emergent complex sociotechnical system that has trust as its core. While responsibility of governance will fall heavily on the corporations using and deploying AI, what corporations do will depend on top-down policy-making by governments and bottom-up demands for trustworthy AI by people. In this regard, AI appears as a call to agency even as the functions of AI challenge the agency of end users. The enforcement of governance at the corporate level can be achieved by ethics-based auditing (Mökander and Axente 2021). Although audits create an added layer of bureaucracy, which could slow down innovation, they can take on different forms based on the scope and context of the project. Indeed, when the credo best defining contemporary techno-solutionism is, "move fast and break things," additional bureaucratic checks may well prove beneficial. Much like universities, which rely on internal review boards to ensure the ethics and integrity of research, corporations could have their internal review boards with different levels of scrutiny, from expedited to full review. Broader scale review boards, such as those like the Food and Drug Administration (FDA), may also be developed to provide external insurance.

9.      **Virtue Ethics**

The ethics of trustworthy AI can be evaluated in relation to three groups of stakeholders: governments, corporations, and people. The conceptual boundaries of trust for each of the three groups of stakeholders and the interdependencies among them can be assessed using three dimensions of trust: ability, integrity, benevolence. Such dimensions should be bolstered by engagement with contemporary research on the nature of human trust and trust propensities for machines, humans, and institutions. Trustworthy AI evolves in tandem with the ethics of AI. A framework of ethics that draws from the three widely recognized ethical approaches – deontological, utilitarian, and virtues – could further improve our trust in AI.

The deontological approach based on rules, and the utilitarian approach based on the cost-benefit calculus of maximizing good and minimizing harm to the largest group of individuals can both inform the development of governance principles. In addition, the virtue-ethics approach (Hagendorff 2022; Vallor 2016) can be a promising addition to nurture the right values in individual developers of AI.

While both the deontological and utilitarian approaches have an important role in macro-level perspectives, virtue ethics, which focuses on the individual, is a burgeoning area that can have an immediate and lasting impact. In other words, top-down ethical principles and actions can serve as extrinsic motivation, and virtue ethics can serve as intrinsic motivation.

Virtue ethics aims to nurture in individuals their own internal compass of values and virtues along with guides to implement them into practice. Virtue ethics inculcates sensibilities like fairness, kindness, compassion, and generosity and examines these virtues in specific contexts relevant to a domain, such as privacy or surveillance. Such investments in training in virtue ethics complemented by ethics-based auditing or internal review processes appear to be promising avenues to build trustworthy AI within the proposed multilevel framework.

## 10.    Conclusion

AI is an exciting technology with enormous risks and benefits. Until recently, the technology has been developed without sufficient consideration of its social implications. Given the pervasive nature of AI and consequential decisions it makes, governance policies are needed. In this chapter we have outlined a framework that emphasizes the interdependencies among three stakeholders – governments, corporations, and citizens. Further, we offer competence, integrity and benevolence dimensions of trust as lenses to examine the interrelationships among the stakeholder groups. We concluded by offering AI ethics as a mechanism to build trust among stakeholders and identify virtue ethics. Along with rule-based ethics, virtue ethics is a promising approach to nurture the right sensibilities required for AI governance.